# Computational Psychology to Embed Emotions into News or Advertisements to Increase Reader Affinity

*Hrishikesh Kulkarni,*   Dr. Prachi Joshi,   Prof (Dr.) Pradip Chande

*Abstract* – Readers take decisions about going through the complete news based on many factors. The emotional impact of the news title on reader is one of the most important factors. Cognitive ergonomics tries to strike the balance between work, product and environment with human needs and capabilities. The utmost need to integrate emotions in the news as well as advertisements cannot be denied. The idea is that news or advertisement should be able to engage the reader on emotional and behavioral platform. While achieving this objective there is need to learn about reader behavior and use computational psychology while presenting as well as writing news or advertisements. This paper based on Machine Learning, tries to map behavior of the reader with the news/advertisements and also provide inputs for affective value for building personalized news or advertisements presentations. The affective value of the news is determined and news artifacts are mapped to reader. The algorithm suggests the most suitable news for readers while understanding emotional traits required for personalization. This work can be used to improve reader satisfaction through embedding emotions in the reading material and prioritizing news presentations. It can be used to map personal reading material range, personalized programs and ranking programs, advertisements with reference to individuals.

## 1.0 INTRODUCTION

Purpose of any news or advertisement is to engage readers or customers. The quality content is silver bullet to it. But is that the only thing. Many quality news get unnoticed while some other news or advertisements touch to heart of readers and customers. Human emotions play a major role in this process. Even there is significant part played by culture, group and associated individuals. While computational sociology tries to model and analyze complex social processes. These social processes and culture of individual also help in deciding the news which can keep reader engage. Understanding computational aspects of human behavior comes under computational psychology and computational anthropology. Research on personalized advertisement and personalized news attracted many researchers. It was always dealing with delivering the highest impact in given small duration. Researcher worked on delivering personalized news and advertisements to television viewers. They clustered TV viewers and mapped advertisement to clusters [1]. Researcher worked on personalized advertisements for digital interactive television [2]. Researchers also worked on personalized news recommendation based on user click behavior [3]. Researchers used log analysis and bayesian recommendation for the same. Researchers worked on adaptive news access so that reader would not be exposed to news not relevant to them [4]. Focus of this work was to reach out to readers interested in the news. Mostly this work was focus on past logs, what person searches and on which type of news he is clicking frequently. Mostly it was based on keyword centric classification of news [5]. The work has further evolved to personality mining. In this movement of personalization researchers got attracted towards mapping human behavior to culture. The work was focused on modeling culture. Researchers even looked at cultural modeling from genomic perspective [6]. Researchers even worked on emotional aspects of culture [7]. There is work reported regarding culture, emotions and depression with reference to behavioral consequences[8]. If we can mine the personality and culture of individuals it can be possible to embed personalized emotions in news or advertisements. The emotions embedding is done with reference to color, shape or even inscribed text. Even words can be selected in headline can have embedded emotions. Embedding emotions and cultural traits in news or advertisements is inspired from embedding emotions into products. When it comes to embedding emotions in products, researcher began with use of colors to map emotions. Kansei was one of the early break through attempt to embed emotions in products and processes [9]. Researchers even worked on use of Kansei engineering in development [10]. Researchers worked on deriving emotions based on expressions of individuals. This expressions are considered in specific scenario to understand context [11].

This paper focuses on embedding emotions in advertisement and news. Advertisement are tend to attach emotionally with the viewer. Simple example is a use of color in advertisement that is more compelling for the viewer or using words those connect him emotionally to the product. Same is true for news. News also have connectivity, words and flow. If it flows so that it can connect to reader it can keep him engage resulting him seeking his attention and achieving the desired impact. In this research readers are classified based on their emotional responses. This helps us to decide emotional theme for that individual. When this theme is associated with the context it is represented with the set of words. This word corpus helps us to identify the news suitable for a particular person. The news are ranked with reference to the candidate. Even chronology can also be decided so that news can be presented to reader in a particular personalized order to produce maximum impact.

## 2.0 PROPOSED METHODOLOGY

The proposed algorithm is based on association of number of vectors. In first part the emotional traits of individuals are determined based on his or her expressions. Here expression come in the form of responses to scenarios. Scenarios or advertisement are represented with parameters impacting customers. In this particular case either same advertisement is presented in different colors or there can be textual slogans associated with the advertisements. While these advertisements are presented to customer he was given options to rank them. Such multiple runs are used for learning and association of his/her emotional traits with advertisement features. Typical features like color, textual inscription and arrangements of different articles are considered. Different colors are associated with different emotions with reference to cultures at different locations. While saffron or orange is associated with religious feelings in Indian continents, white is associated with piece. Similarly some words specifically depicts personalized emotions with reference to context. The algorithm selects the words those can deliver highest impacts on customers increasing his affinity towards the product. Similarly news headlines are tuned in personalized way to deliver grater impact on readers. Every word in advertisement produce an unique emotional impact on reader. This impact either increase his affinity towards that particular news or product or some times it can repel him away from that product. In case the news can establish emotional affinity with the reader it leads to a situation where he reads that news carefully and even start following news by same presenter, same channel or even in the same topic. This is even true for advetisements. If advertisements can establish emotional bonding with the reader he/she start looking for the products in the advertised. This leads to emotional affinity of a person with a particular brand. This can result in person buying many other products by same brand. This emotional affinity sometimes overrides his personal preferences and utility parameters. Even he starts perceiving the product as one of the most useful products and becomes a sort of brand ambassador of that particular brand.

## 3.0 EMBEDDING EMOTIONS

How to embed emotions in the advertisements or news still remains a challenge in front of researchers. To embed emotions we need to identify personalized emotions. With reference to these personalized emotions we have to map perception and cognition so that when customer will look at advertisement he should feel an emotional attachment with the product. In case of news or advertisement the corpus of words mapping to the emotions of customer is used. The most suitable word is determined based on personality vector. To achieve the desired emotional impact three major factors are considered. These three factors are color, shape and presentation. In case of presentation order of images as well as inscribed text is also considered. Presentation includes other additional things like order in presentation and inscribed text.

During the first step the personality mining is used to formulate personality vector. Based on customers selections of answers in the form of visual objects personality vector is derived. The associations among customer expressions are used for the same.

Hypothesis 1: Expression based personality vector can be used to predict emotional affinity of customer with the news or advertisement.

Hypothesis 2: Color, shape and presentation along with inscribed text can be used to embed emotions in news and advertisements

Eq. 1 gives a personality vector while eq. 2 gives emotion vector of a candidate.

$$PV = \begin{bmatrix} PV_1 \\ PV_2 \\ PV_3 \\ \vdots \\ PV_n \end{bmatrix} \qquad (1)$$

$$EV = \begin{bmatrix} EV_1 \\ EV_2 \\ EV_3 \\ \vdots \\ EV_n \end{bmatrix} \qquad (2)$$

Same advertisement with images is presented in multiple colors, different inscribed texts and changing background to capture emotions. In similar fashion total five advertisements are presented to customer. The same exercise is conducted with news articles. In this case headline are changed with some specific set of words. The responses by readers/customers are captured. It is observed that particular percept results in a sort of emotions. Thus it is about identifying perceptions like color, shape, inscribed text and size so that it impacts positively on emotions of the customer. It could result in emotional affinity. The emotional response of the customers is measured through the actions and expressions. The proposed method takes this complete

emotional need identification and embedding emotions in the advertisements/news as a four step process:

1. Understanding customer's affective needs: (this is a learning phase of the algorithm.) While doing this, customer is presented a series of advertisements or news of his interest. The variations in affective components impacting percept help us to identify his/her affective needs. An experiment is conducted for 50 candidates and five advertisements. For each product the variations are applied.
2. In case of new candidates based on their personal data the affective needs are calibrated. With reference to affective needs customer is presented news or advertisements. In case of news even series of news are also considered. Part of data which is already labeled is used for testing.
3. Finally, the representative emotional vectors and personality vectors are used to map a customer to the most suitable advertisement or news.
4. In this case corpus based on customers' emotional traits the affective components and features selected for embedding emotions.

Any new candidate is classified in the emotional range form 1 to 5. For the convenience of computation we have take n from 1 to five but it can very easily be extended to bigger number provided we get comparable accuracy.

## 4.0 RESULTS

Emotions are embedded in the any object in different ways. In case of news it can be in the form of words those depict personalized emotions those can keep reader to go in details and read the next version of it. It could be in the form of percept in case of advertisements. It could be shade or color or changes in color, it could be words the way they appear, it could be words in the title of news or it could even be order in which news are presented. As discussed in previous section we have clustered the candidates based on their responses in different emotional zones. Personality vectors and emotion vectors are used for classification. We have collected response expressions for 500 candidates. We clustered them using affinity index.

**Table I: Emotions embedded through text in title of news**

| Candidate | Context One | | | | |
|---|---|---|---|---|---|
| | *Cluster One* | *Cluster Two* | *Cluster Three* | *Cluster Four* | *Cluster Five* |
| Cluster I | | | | | |
| 1 | 3 | 1 | 3 | 1 | 1 |
| 2 | 3 | 0 | 4 | 2 | 2 |
| 3 | 2 | 0 | 3 | 1 | 1 |
| 4 | 2 | 1 | 4 | 3 | 2 |
| 5 | 2 | 2 | 4 | 5 | 4 |
| Cluster II | | | | | |
| 6 | 2 | 1 | 2 | 4 | 4 |
| 7 | 4 | 0 | 3 | 3 | 3 |
| 8 | 1 | 2 | 2 | 3 | 4 |
| 9 | 2 | 4 | 3 | 1 | 3 |
| 10 | 4 | 2 | 3 | 2 | 2 |

Inclination toward reading the similar news or news in that series in the area of interest of the candidate is calibrated on scale 0 to 4. The news selected in such a way that – it is a news he would love to read. 0 indicates complete disagreement towards reading similar news while 4 means reader is desperate towards reading similar news.

Data set of 10 candidates for different word clusters is given in Table 1. The same way data is collected for other emotional features. These clusters are used for learning. For test set data, personality vector is used for mapping emotional traits of individuals. Then based on cluster and other personality analysis of individual a product with mapping emotional features is presented to reader. Responses for the other combinations are also collected from customers.

Table 2 gives comparison between actual ranks and expected ranks for given set of news. In 60% of cases the embedded emotions in the form of text clusters as per emotional traits of reader and the news is ranked as best one. In 20% of the cases there is another news reader ranks above this recommended news and in rest of the cases there are two or more news articles out of set of five news articles, reader ranks above this recommended news article.

**Table II: comparison with skill based classification**

| Serial No | Scenario One | |
|---|---|---|
| | *Expected Rank* | *Actual Ranks* |
| 1 | 1 | 2 |
| 2 | 1 | 1 |
| 3 | 1 | 1 |
| 4 | 1 | 1 |
| 5 | 1 | 3 |
| 6 | 1 | 2 |
| 7 | 1 | 1 |
| 8 | 1 | 1 |
| 9 | 1 | 2 |
| 10 | 1 | 1 |

a. All the results are compared with already labeled data in test set

Table III depict percentage accuracy with reference to embedded emotions for five classes of different emotional types.

**Table III: Percentage Accuracy**

| Emotional Type (Class) | Context |
|---|---|
| | *Affinity Index* |
| 0 | 61 |
| 1 | 61 |
| 2 | 67 |
| 3 | 72 |
| 4 | 70 |

Fig. 1 depict class wise percentage accuracy. It clearly suggests that there is higher accuracy observed in classes 3 and 4 as compare to class 1, 2 and 3. Fig 2 depicts comparison between actual and expected rank of emotion embedded products. It can be observed from the diagram that for six out of 10 samples the rank is matching exactly. The representative rank comparison makes sure that embedding emotions using affinity index can make products more appealing to customers.

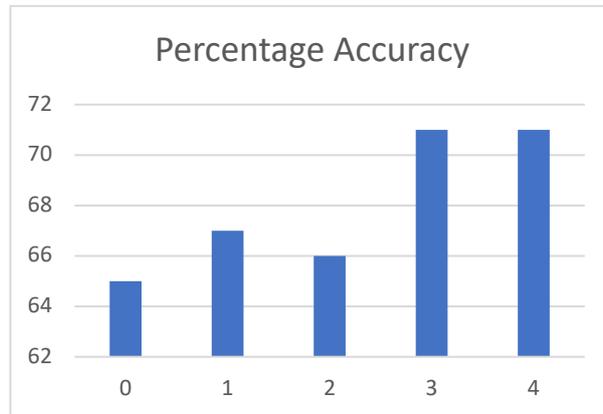

Fig. 1  Classwise accuracy

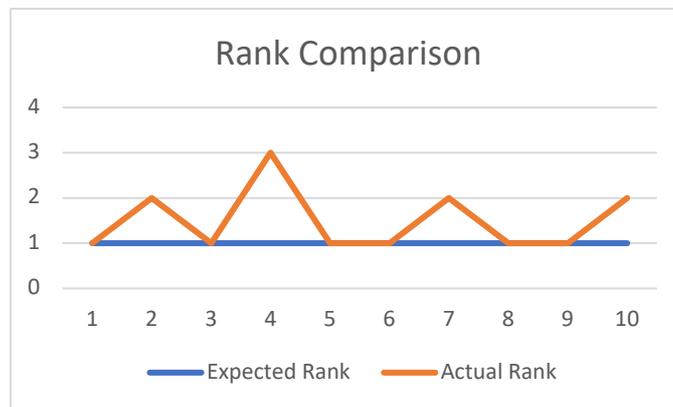

Fig. 2  Comaprison of Results

## 5.0 CONCLUSION

It's always challenging to embed emotions into any object, news or advertisement. Color and Text have always been choice of researchers for their experiments. While Kansei Engineering focuses on building personalized products and have emotions embedded in the process. Computationally how to measure emotions? is challenging and hence only way is to collect response from customer or track his emotions through expressions. Computational psychology experiments were based on color selection. In this paper, we have proposed a method for embedding emotions into the advertisements and news articles and experiments to measure the success and impact of it on readers and customers. The association and affinity index is used for mapping individuals to emotion vectors. The results are measured based on ranking done by readers and ranking done by Affinity Based Association. In most of the case the algorithm confirmed accuracy greater than 62%. This method can be enhanced to associate more than one advertisements, deciding sequence of news to achieve desired impact and also to attract readers.

## 6.0 REFERENCES

[1] G. Lekakos , G Giaglis, Delivering Personalized Advertisements in Digital Television: A Methodology and Empirical evaluation,
 https://pdfs.semanticscholar.org/42a3/ea54f42532569f67831c8ff0c3e167c931b7.pdf

[2] Pramataris K., Papakyriakopoulos D., Lekakos G., Mylonopoulos N.: Personalized Interactive TV Advertising: The IMEDIA Business Model. Journal of Electronic Markets, Volume 11 (1): 1– 9. (2001).

[3] Jiahui Liu, Peter Dolan, Elin Rønby Pedersen, Personalized News Recommendation Based on Click Behavior,
 https://static.googleusercontent.com/media/research.google.com/en//pubs/archive/35599.pdf

[4] Billsus, D., Pazzani, M. J., User Modeling for Adaptive News Access, User Modeling and User-Adapted Interaction, v.10 n.2-3, p.147-180, 2000

[5] Carreira, R., Crato, J. M., Gonsalves, D., Jorge, J. A. Evaluating adaptive user profiles for news classification, Proceedings of the 9th international conference on Intelligent user interfaces, 2004

[6] Contextual Geometric Structures: modeling the fundamental components of cultural behavior. Proceedings of Artificial Life, 13, 147-154 (2012)

[7] Chentsova-Dutton, Y.E. & Tsai, J.L. (2010). Self-focused attention and emotional reactivity: The role of culture. Journal of Personality and Social Psychology, 98, 507-519.

[8] Tsai, J.L. (2007). Ideal affect: Cultural causes and behavioral consequences. Perspectives on Psychological Science, 2, 242-259.

[9] Lee, S.H., Harada, A., Stappers, P.J., 2002, Pleasure with Products: Design based on Kansei, published in Green, W. and Jordan, P., "Pleasure with Products: Beyond usability" ed. Taylor & Francis, London, p. 219-229

[10] Schütte, S., 2005, Engineering Emotional Values in Product Design – Kansei Engineering in Development, Doctoral Thesis, Institute of Technology, Linköping Universitet, Sweden. ISBN: 91-85299-46-4.

[11] Hrishikesh Kulkarni, "Intelligent Context Based Prediction using Probabilistic Intent-Action Ontology and Tone Matching Algorithm", IEEE Conference, ICACCI, Manipal, Oct 2017 (https://ieeexplore.ieee.org/document/8125916/)